\documentclass[final, 11pt]{article}

\usepackage{ltexpprt} \usepackage{amsmath} \usepackage{amssymb} \usepackage{colordvi} \usepackage{color} \usepackage{epsf} \usepackage{pst-grad} \usepackage{pifont} \usepackage{mathcomp} \usepackage{epsfig} \usepackage{fancyhdr} \usepackage{graphics} \usepackage{wasysym} \usepackage{algorithm} \usepackage{algorithmic} \usepackage{multirow}

  \newtheorem{remark}{Remark} \newtheorem{example}{Example} \newtheorem{definition}{Definition}

\begin{document}

\title{Algorithms for Scheduling Weighted Packets with Deadlines in a Bounded Queue}

\author{Fei Li\thanks{Department of Computer Science, George Mason University. {\tt lifei@cs.gmu.edu.} Part of this work appears in the Proceedings of the 28th IEEE International Conference on Computer Communications (INFOCOM 2009)~\cite{L09b}.}}

\maketitle

\pagestyle{plain}


\begin{abstract}

Motivated by the Quality-of-Service (QoS) buffer management problem, we consider online scheduling of packets with hard deadlines in a finite capacity queue. At any time, a queue can store at most $b \in \mathbb Z^+$ packets. Packets arrive over time. Each packet is associated with a non-negative value and an integer deadline. In each time step, only one packet is allowed to be sent. Our objective is to maximize the total value gained by the packets sent by their deadlines in an online manner. Due to the Internet traffic's chaotic characteristics, no stochastic assumptions are made on the packet input sequences. This model is called a {\em finite-queue model}.

We use competitive analysis to measure an online algorithm's performance versus an unrealizable optimal offline algorithm who constructs the worst possible input based on the knowledge of the online algorithm. For the finite-queue model, we first present a deterministic $3$-competitive memoryless online algorithm. Then, we give a randomized ($\phi^2 = ((1 + \sqrt{5}) / 2)^2 \approx 2.618$)-competitive memoryless online algorithm.

The algorithmic framework and its theoretical analysis include several interesting features. First, our algorithms use (possibly) modified characteristics of packets; these characteristics may not be same as those specified in the input sequence. Second, our analysis method is different from the classical potential function approach. We use a simple charging scheme, which depends on a clever modification (during the course of the algorithm) on the packets in the queue of the optimal offline algorithm. We then prove that a set of invariants holds at the end of each time step. Finally, we analyze the two proposed algorithm in a relaxed model, in which packets have no hard deadlines but an order. We conclude that both algorithms have the same competitive ratios in the relaxed model.

\end{abstract}


\section{Introduction}

In the last three decades, routers in the Internet continue supporting more and more applications.  Currently, most routers forward packets in a First-In-First-Out ({\tt FIFO}) manner and treat all packets equally. However, the diversity of applications has resulted in heterogeneity and unpredictable network traffic. Thus, it is more reasonable to consider differentiation among packets from different types of applications (see \cite{WP98, AMRR00, KLMPSS04, AMRR05} and the references therein).  For instance, we could specify values for packets to represent their priorities. Also, we may like to assign hard deadlines to packets in time-critical applications. These concerns have made buffer management at routers significant in providing effective quality of service (QoS) to various applications.

\begin{figure}[htp]
\centering
    \includegraphics[width=.4\textwidth]{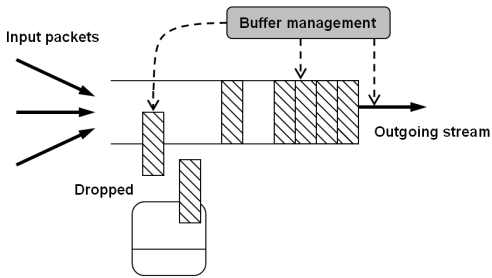}
    \caption{Buffer management is in charge of processing arriving packets from the input streams and delivering packets out of the buffer as outgoing streams.}
\label{fig:buffer}
\end{figure}

One kind theoretical research on QoS buffer management starts from three paper by Aiello et al.~\cite{AMRR00}, by Kesselman et al.~\cite{KLMPSS04} and by Hajak~\cite{H01}, where a model called a {\em bounded-delay model} is proposed.  In this model, time is discrete. Packets arrive over time, and they are buffered upon arrivals. The queue capacity is unlimited. An arriving packet $p$ has a non-negative {\em value} $w_p \in \mathbb R^+$ and an integer {\em deadline} $d_p \in \mathbb Z^+$ by which it should be transmitted; after $d_p$, $p$ expires.  In each time step, at most one packet can be sent. The objective is to maximize the {\em weighted throughput}, which is defined as the total value of the transmitted packets by their deadlines. Fig.~\ref{fig:buffer} illustrates the functionalities of the online buffer management algorithms, which process newly arriving packets and send one packet out of the buffer in each time step. The buffer size and deadlines of packets limit the number of {\em pending packets}\footnote{A pending packet is a packet in the buffer whose deadline has not expired yet. For a given time step, every pending packet is eligible for sending.} in the queue.

Realizing that the capacity of a queue buffering packets is limited and such queue is a shared resource for multiplexing packets inside routers, Azar and Levy extend the single buffer bounded-delay model to multiple buffers. They consider scheduling packets with deadlines in multiple finite capacity buffers, all of which have the same finite capacities~\cite{AL06}. In this paper, we study the single queue scheduling problem, in which the capacity of the queue is finite.

In the ideal case, if the release time, value, and deadline of each packet are known ahead of time, an optimal schedule can be found efficiently; we call this the {\em optimal offline algorithm}. For instance, given no constraint over the queue capacity, the the optimal schedule can be found by computing a maximum weighted matching on a convex bipartite graph.  However, we do not know all such information ahead of time. Rather, packets arrive {\em online}, and we only learn about a packet and its associated characteristics when it actually arrives. Furthermore, without a realistic model of the network traffic~\cite{WP98, FP01}, we cannot achieve good stochastic performance guarantee. Hence, we study the worst-case analysis for algorithms in queueing packets, without any assumptions over the input sequence.

Competitiveness has been widely accepted as the metric to measure an online algorithm's worst-case performance in theoretical computer science~\cite{BY98}. In this paper, we design and analyze better deterministic and randomized online algorithms, in terms of competitive ratio, for scheduling packets in a finite capacity queue. A deterministic (randomized) online algorithm ${\tt ON_d}$ (${\tt ON}_r$) is called {\em $k$-competitive} if its (expected) weighted throughput on {\em any} instance is at least $1 / k$ of the weighted throughput of an optimal offline algorithm on this instance:
\begin{eqnarray*}
k & = & \max_{\cal I} \frac{{\tt OPT}({\cal I}) - \alpha}{{\tt ON_d}({\cal I})}, \ \ \ \mbox{${\tt ON}_d$ is a deterministic algorithm} \\
k & = & \max_{\cal I} \frac{{\tt OPT}({\cal I}) - \alpha}{{\bf E}[{\tt ON_r}({\cal I}, \ r)]}, \ \ \ \mbox{${\tt ON}_r$ is a randomized algorithm}
\end{eqnarray*}
where $\alpha$ is a constant, ${\tt OPT}({\cal I})$ is the optimal solution of an input $\cal I$, and $r$ is the set of random variables flipped by a randomized online algorithm ${\tt ON}_r$. The parameter $k$ is known as the online algorithm's {\em competitive ratio}~\cite{BY98}~\footnote{In real-time scheduling terminologies, $1 / k$, the reciprocal of the competitive ratio, is called {\em competitive factor}. We use the term {\em competitive ratio}, which is widely recognized in the area of online computation.}. If $k$ is not a constant, such an algorithm with competitive ratio $k$ is called {\em non-competitive}. If the additive constant $\alpha \le 0$, the algorithm ${\tt ON}_d$ (${\tt ON}_r$) is called {\em strictly $k$-competitive}. Note that for a randomized algorithm, the role of randomization is purely internal to the randomized algorithm. No stochastic assumption is made on the input. Also note that the optimal offline algorithm is called the {\em adversary} of the online algorithm since the input sequence constructed by the offline optimal algorithm is allowed to maximize the competitive ratio $k$.


\subsection{Problem setting.}

The model we study here is called the {\em finite queue model}, in which we consider scheduling packets with deadlines in a finite capacity queue. In our model, time is discrete. Packets arrive over time and each packet $p$ is associated with a non-negative weight $w_p \in \mathbb R^+$ and an integer deadline $d_p \in \mathbb Z^+$.  $d_p$ specifies the time by which $p$ should be sent. If $p$ is transmitted by its deadline $d_p$, $p$ contributes our objective by a value $w_p$.  (We use ``value'' and ``weight'', ``queue'' and ``buffer'' interchangeably.) This model is preemptive, which means that packets already existing in the queues can be dropped at any time before they are served. If a packet is dropped, this packet cannot be delivered any more.

There is only one queue with a limited capacity of $b \in \mathbb Z^+$. At any time, the queue can store no more than $b$ packets. Packets that expire are dropped immediately. At most one packet can be sent in each time step. Our target is to maximize the total value of the packets sent by their deadlines. Remember that in a bounded delay model~\cite{KLMPSS04, H01}, the queue capacity is infinite. The finite queue model generalizes the bounded delay model: if the queue capacity is larger than any packet's {\em slack time}, which is defined as the difference between the packet's deadline and its release time, the finite queue model is the bounded-delay model.


\subsection{Related work.}

Since the first QoS buffer management model (the bounded-delay model) was introduced in~\cite{AMRR00, AMRR05}, many researchers have considered this model as well as its variants~\cite{KLMPSS04, H01, CF03, CCFJST06, CJST07, LSS05, LSS07, EW07}. Most of these studies consider the single queue case, which does not have an explicit limit over the queue capacity. The best known lower bound of competitive ratio of deterministic algorithms is $\phi = (1 + \sqrt{5}) / 2 \approx 1.618$~\cite{H01, CF03, AMZ03}; this lower-bound also applies to instances in which the deadlines of the packets (weakly) increase with their release dates.

For an arbitrary deadline instance, a simple greedy algorithm that always schedules a maximum-value packet in the queue is $2$-competitive~\cite{H01, KLMPSS04}.  A generalization of the greedy algorithm, called ${\tt EDF}_{\alpha}$, which schedules the earliest packet with a value at least $1 / \alpha$ ($\alpha \ge 1$) of the maximum-value of a packet~\cite{CCFJST06}, has a competitive ratio of (asymptotically) $2$.  Chrobak et al.~\cite{CJST07} discuss a clever modification that results in an algorithm with a competitive ratio of $64 / 33 \approx 1.939$. This algorithm employs a status bit to help schedule packets and hence, it is not memoryless~\footnote{An algorithm is called {\em memoryless} if this  algorithm makes its scheduling decision only based on the packets in the current queue but not on the historical information.}.

For instances in which the deadlines of the packets (weakly) increase with their release dates, Li et al.~\cite{LSS05} propose an optimal deterministic online algorithm {\tt MG} whose competitive ratio is $\phi$. Unfortunately, this improved competitive ratio is achieved by exploiting the deadline assumption; on general instances, {\tt MG}, like the greedy algorithm and ${\tt EDF}_{\alpha}$, is also $2$-competitive.  Applying similar analysis approach, but in a more complicated way, Li et al. provide a $3 / \phi \approx 1.854$-competitive deterministic algorithm~\cite{LSS05} for the general model. Independently, Englert and Westermann present a $1.894$-competitive deterministic memoryless algorithm and a ($2 \sqrt{2} - 1 \approx 1.828$)-competitive deterministic algorithm~\cite{EW07}, which is not memoryless. Closing the gap of $[1.618, \ 1.828]$ between the lower bounds and the upper bounds of competitive ratio of deterministic online algorithms is still a difficult open problem. Randomization on the bounded-delay model is considered in~\cite{CCFJST06}. A randomized online algorithm with a competitive ratio of $e / (e - 1) \approx 1.582$ is proposed. The lower bound of competitive ratio of randomized algorithms is $1.25$. How to tighten the gap of $[1.25, \ 1.582]$ in the randomized model still remains open.

Azar and Levy consider the multi-buffer model in which multiple queues are bounded in their capacities and packets can have arbitrary deadlines. Notice that if the queues are unlimited in capacities, this model is the same as the bounded-delay single queue model. The lower bound $1.618$ for the bounded-delay model directly applies on the multi-queue model. In~\cite{AL06}, the authors give a deterministic memoryless $9.82$-competitive algorithm. In this paper, we improve the lower bound to $2$ for a family of deterministic online algorithms (this lower bound also applies to the finite queue model). To our knowledge, there is no published work of randomized algorithms on the multi-buffer model. Also, our proposed single finite queue model, which is more realistic for buffer management, has not been addressed in recent literature. For the single queue case, the competitive ratio of the algorithm in~\cite{AL06} $9.82$ still applies.

There has also been work on another model in which the queue capacity is bounded. In this model, packets have no deadlines but weights, and the {\tt FIFO} discipline is enforced in delivering packets~\cite{LP02, KMS05, BFKMSS04} --- packets should be sent in the same order as they arrive.  Some researchers also consider packet scheduling in multiple {\tt FIFO} input queues connecting one output queue~\cite{AR03, AR04, AS05, IT06}:  Every queue obeys the {\tt FIFO} constraint in delivering weighted packets and each arriving packet has only one destined queue.


\subsection{Our contributions.}

This paper provides theoretical bounds for algorithms on the finite queue model, which considers the finite capacity constraint for buffer management under a more practical modeling. Our main contributions include
\begin{enumerate}

\item A strictly $3$-competitive deterministic memoryless online algorithm {\tt ME} for the finite queue model (in Section~\ref{sec:me}).

\item A strictly ($\phi^2 = ((1 + \sqrt{5}) / 2)^2 \approx 2.618$)-competitive randomized memoryless online algorithm {\tt RME} for the finite queue model (in Section~\ref{sec:rme}).

\item A new analysis method including a charging scheme and a set of invariants.

\end{enumerate}

Table~\ref{tbl:results} summaries the competitive ratios of those known algorithms for the bounded-delay model, its variants, and our results on the finite queue model.

\begin{table*}[ht]
\begin{center}
\begin{tabular}{|p{4cm}|p{6cm}|p{6cm}|}

\hline

Models & Upper bound of competitive ratio & Lower bound of competitive ratio\\ \hline \hline

General bounded-delay model & Deterministic algorithms: $1.854$~\cite{LSS07} $1.828$~\cite{EW07} & Deterministic algorithms: $1.618$~\cite{KLMPSS04, H01} \\

& Randomized algorithms: $e / (e - 1) \approx 1.582$~\cite{CCFJST06} & Randomized algorithms: 1.25~\cite{CCFJST06} \\ \hline

Agreeable deadline bounded-delay model & Deterministic algorithms: $1.618$~\cite{LSS05} & Deterministic algorithms: $1.618$~\cite{KLMPSS04, H01} \\ \hline

General finite queue model & Deterministic algorithms: $3$ (in this paper) & A broad family of deterministic algorithms: $2$ (in this paper) \\

& Randomized algorithms: $2.618$ (in this paper) & - \\ \hline

\end{tabular}

\end{center}

\caption{Summary of competitive ratios for the bounded-delay model and the finite queue model. Upper bounds are achieved by some known algorithms. Any online algorithm cannot achieve a competitive ratio less than the lower bound.}

\label{tbl:results}

\end{table*}

To supplement our work on the finite queue model, we also provide an optimal offline algorithm (in Section~\ref{sec:offline}).


\section{Algorithm {\tt ME} and Its Analysis}
\label{sec:me}

In this section, we introduce a deterministic memoryless online algorithm {\tt ME} to schedule packets with deadlines in a single finite capacity queue. {\tt ME} stands for ``Modified {\tt EDF}''. We first discuss the intuitions behind the algorithm {\tt ME}. Then, we present {\tt ME} and its analysis.


\subsection{Intuitions of designing {\tt ME}.}

We commence our study at a well-known real-time scheduling algorithm called {\tt EDF} (``Earliest-Deadline-First''). {\tt EDF} is one of the most important (and ever analyzed) dynamic priority algorithm, and the priority of a job (or a packet) is inversely proportional to its absolute
deadline. In each time step, {\tt EDF} schedules the packet with the earliest deadline~\cite{CCFJST06}. The following example shows that even {\tt EDF} calculates the best schedule sequence among all pending packets, it does not have a constant competitive ratio for the finite queue model~\footnote{Such an algorithm is called {\em non-competitive}.}. In this example, the queue capacity is $b \in \mathbb Z^+$ and we use $(w, \ d)$ to represent a packet with value $w$ and deadline $d$. We use $d = \infty$ to denote a packet with a very large deadline.

\begin{example}
{\em Initially, the algorithm's queue is empty. In the first time step, $b - 1$ packets of $(\epsilon, \ i)$, $i \ = \ 1, \ 2, \ \ldots, \ b - 1$ and one packet $(1, \ \infty)$ arrive. In each of the following $b - 2$ time steps $2, \ \ldots, \ b - 1$, only one packet $(1, \ \infty)$ is released to the queue.

When a new packet arrives, {\tt EDF} tries to accept it and drops the minimum-value packet only if the queue is overflow. {\tt EDF} sends the earliest-deadline packet in each time step. In our instance, {\tt EDF} sends the packet $(\epsilon, \ i)$ in each time step $i \ = \ 1, \ 2, \ \ldots, \ b - 1$ and all released packets $(1, \ \infty)$ are stored in its queue till the end of step $b - 1$. On the contrary, the optimal offline algorithm sends one packet $(1, \ \infty)$ in each of the first $b - 1$ time steps, and only one packet $(1, \ \infty)$ remains in its queue at the end of step $b - 1$.

At the beginning of step $b$, {\tt EDF}'s queue is full and has $b$ packets $(1, \infty)$. Now $b$ packets of $(1 - \epsilon, 2 \cdot b)$ arrive but since they have weights smaller than the packets already in the buffer, they are dropped. Notice that {\tt EDF} has lost $b - 1$ packets of $(1 - \epsilon, \ 2 \cdot b)$ due to the overflow happening in step $b$. At the beginning of each step $b + 1, \ b + 2, \ \ldots, \ 2 \cdot b$, one packet $(\epsilon, \ b + i)$ ($i \ = \ 1, \ 2, \ \ldots, \ b$) arrives. Given that all packets already in the queue keep their deadlines $\infty$, {\tt EDF} sends packet $(\epsilon, \ b + i)$ in each time step. On the contrary, the offline optimal algorithm sends $(1 - \epsilon, \ 2 \cdot b)$ in each step.

At the beginning of step $2 \cdot b + 1$, $b$ packets of $(1 - \epsilon, \ 3 \cdot b)$ arrive and {\tt EDF}'s queue has a packet $(1 - \epsilon, \ 3 \cdot b)$ and $b - 1$ packets $(1, \ \infty)$. {\tt EDF} sends a packet $(1 - \epsilon, \ 3 \cdot b)$ in step $b$. Notice that {\tt EDF} has lost $b - 1$ packets of $(1 - \epsilon, \ 3 \cdot b)$ due to the overflow happening in step $2 \cdot b + 1$. At the beginning of each time step $2 \cdot b + 2, \ 2 \cdot b + 3, \ \ldots, \ 3 \cdot b$, one packet $(\epsilon, \ 2 \cdot b + i)$ ($i \ = \ 2, \ 3, \ \ldots, \ b$) arrives. Given that all packets already in the queue have deadline $\infty$, {\tt EDF} sends the packet $(\epsilon, \ 2 \cdot b + i)$ and keeps those in the queue without worrying about them being expired, hoping to send them in the future. On the contrary, the offline optimal algorithm sends $(1, \ \infty)$ in each time step.

We repeat this pattern. In the interval between $2$ overflows (a period of $b$ time steps), the optimal offline algorithm sends $b$ large-value packets with value $1 - \epsilon$ and {\tt EDF} sends only one large-value packet and $b - 1$ small value packets with value $\epsilon$. We find that {\tt EDF} cannot achieve a total value more than $1 / b$ of what an offline optimal algorithm does. Assume we run $n$ rounds of the same pattern of released packets. {\tt EDF} sends all packets $(1, \ \infty)$ after $n$ rounds. The competitive ratio $c$ of {\tt EDF} is (assume $n \cdot \epsilon = 1$)
\begin{displaymath}
c = \frac{1 \cdot b + (1 - \epsilon) \cdot b \cdot n + \epsilon \cdot b}{[\epsilon \cdot (b - 1) + 1] \cdot n + 1 \cdot b} \ge \frac{b}{1 + (2 \cdot b - 1) / n} \ge b - \frac{1}{1 + n / (2 \cdot b - 1)}.
\end{displaymath}

Given $b$ is large and if we repeat above pattern for at least $1 / \epsilon$ times, {\tt EDF} is not competitive in scheduling packets with deadlines in the finite capacity queue because the competitive $c$ is not bounded by a constant.} $\square$
\label{ex:edfex}
\end{example}

Example~\ref{ex:edfex} reveals that {\em even {\tt EDF} keeps the set of pending packets with the maximum total value in each time step, it is not competitive}. The underlying idea of using {\tt EDF} is that we do not drop any packet $p$ unless $p$ is going to expire at time $d_p$ or in the queue, there are more packets with no less value than $w_p$ having to be sent before $d_p$. The non-competitiveness of {\tt EDF} over the above instance implies that we need a better method to identify whether a more valuable packet should be sent even well before its ``real deadline''. For example, packet $(1 - \epsilon, \ \infty)$ released in step $b + 1$ in Example~\ref{ex:edfex} should be sent early instead of being overflowed by later packets. Thus, it is critical for us to define and associate a ``virtual deadline'' with each packet, instead of the real deadline assigned, to denote the ``best latest time'' by which a packet should be sent. Inspired by the {\tt EDF} instance in Example~\ref{ex:edfex}, we propose an algorithm {\tt ME}, which keeps track of a packet's importance with respective to the others by using its ``virtual deadline''.


\subsection{Algorithm {\tt ME}.}

At first, we introduce some notation. A packet $p$ arrives at an integer time $r_p \in \mathbb Z^+$. $p$ has a non-negative value $w_p \in \mathbb R^+$ and an integer deadline $d_p \in \mathbb Z^+$. Given a time $t$, we denote the buffer of an algorithm {\tt A} to be $Q^{\tt A}_t$. All buffer slots in $Q^{\tt A}_t$ are indexed as $0, \ 1, \ \ldots, b - 1$. We use $Q^{\tt A}_t(i)$ to denote the packet in the buffer slot indexed as $i$. If there is no packet in a buffer slot $i$, $Q^{\tt A}_t(i)$ is a {\em null packet}. We associate each packet $p$ a {\em virtual deadline} $t_p$. At $p$'s arrival, $t_p$ is initialized as its real deadline specified by the adversary in the input sequence.

Given a set of pending packets, a {\em provisional schedule} specifies which packet should be sent in which time step no later than its {\em virtual deadline}, assuming no future arrivals. Given a set of pending packets with virtual deadlines, an {\em optimal provisional schedule} is the one that achieves the maximum total value of packets among all provisional schedules on pending packets. The optimal provisional schedule gives a greedily optimal schedule of all the pending packets at time $t$: If there is no future arrivals, sending the packets each in one time step following the optimal provisional schedule is optimal for maximizing the total gain.

We develop our algorithm {\tt ME} from our considerations on the {\tt EDF} instance in Example~\ref{ex:edfex}. {\tt ME} consists of $3$ parts:
\begin{enumerate}

\item Based on the virtual deadlines of packets, calculate the optimal provisional schedule of sending the pending packets in the queue (including the new arrival in this time step), assuming there is no future arrivals (see Algorithm~\ref{alg:ops}).

\item Update the virtual deadlines of packets, if needed (see Algorithm~\ref{alg:me}).

\item Send the packet with the earliest virtual deadline {\em or} the maximum-value packet, based on the ratio of these two packets (see Algorithm~\ref{alg:me}).

\end{enumerate}

Assume for each packet $p$, we have known its virtual deadline $t_p$. The following procedure {\tt OPS} ({\tt OPS} stands for ``Optimal Provisional Schedule'') greedily calculates the optimal provisional schedule from the set of pending packets ${\bf S}$ at time $t$. In {\tt OPS}, we first sort packets in non-increasing weight order. Then we pick up a packet $p$ and put it into an empty queue as later as we could. If we cannot find such an empty buffer slot for $p$, this packet is discarded. All packets selected to be put into the queue are claimed to be in the optimal provisional schedule. {\tt OPS} is described in Algorithm~\ref{alg:ops}.

\begin{algorithm}
\caption{{\tt OPS}({\bf  S}, \ t)}
\begin{algorithmic}[1]

\STATE Sort all packets in $\bf S$ in non-increasing weight order, with ties broken in favor of the larger virtual deadlines.

\WHILE{${\bf S} \neq \emptyset$}

\STATE Pick up a packet $p$ from $\bf S$.

\FOR{each buffer slot $i$ indexed from $\min\{t_p - t, \ b - 1\}$ down to $0$}

\IF{there is no packet in the buffer slot indexed as $i$}

\STATE Put $p$ into the $i$-th buffer slot.

\STATE Remove $p$ from $\bf S$.

\STATE {\bf Break}.

\ENDIF

\ENDFOR

\IF{$p$ is not added into the queue}

\STATE Discard $p$.

\ENDIF

\ENDWHILE

\STATE Sort all packets in the queue in non-decreasing virtual deadline order, with ties broken in favor of the larger value packets.

\end{algorithmic}

\label{alg:ops}

\end{algorithm}

Using an interchange argument, we prove the optimality of {\tt OPS}.
\begin{lemma}
{\tt OPS}(${\bf S}$) calculates the optimal provisional schedule for a set of pending packets $\bf S$.
\label{lemma:ops}
\end{lemma}

\begin{proof}
In the algorithm {\tt OPS}, for each packet $p$, we either move $p$ into the queue or we permanently discard it. We finalize the provisional schedule $\tilde S$ in a greedy manner. To prove Lemma~\ref{lemma:ops}, it is sufficient to prove that for {\em each} packet $p$ in the optimal provisional schedule $S^*$, $\tilde S$ and $S^*$ choose the same set of packets to put into buffer slots $[\tilde S(p) - t, \ b - 1]$, where $\tilde S(p)$ denotes the time step in which $p$ is put into $\tilde S(p)$, given the set of pending packets and the assumption of no future arrivals. Without loss of generality, we assume all packets in $S^*$ are sorted in non-decreasing deadline order, with ties broken in favor of the larger value packets.

We assume there exists an optimal provisional schedule $S^*$. If ${\tilde S} = S^*$, Lemma~\ref{lemma:ops} holds immediately. Let us assume ${\tilde S} \neq S^*$.  We then compare the packets scheduled in $\tilde S$ and $S^*$ from the buffer slot indexed as $b - 1$ in reverse order. $q$ is the first packet appearing in the schedule $S^*$ that is different from its counterpart in $\tilde S$ (in the backward manner) and the corresponding time slot in $\tilde S$ contains $p$. A packet $p \neq q$ must be found.

We apply the interchange argument to prove ${\tilde S} = S^*$. From our procedure of selecting packets in {\tt OPS}, we know that any packet in a queue from $\tilde S(p)$ to its virtual deadline $t_p$ has a value larger than or equal to $w_p$.

\begin{enumerate}

\item If $w_q > w_p$, then, $q$ should be chosen before $p$ when we create $\tilde S$ and $q$ should be put in the position $\tilde S(p)$ instead of the position of $p$.

\item If $w_q \le w_p$, the optimal provisional schedule $S^*$ should contain $p$ since it includes $q$ ($t_p$ and $t_q$ are not before the time slot $S^*$ schedules $q$). Without losing any value, $S^*$ can swap $p$ and $q$ since $p$ is not in any buffer slot from $S^*(p)$ to $t_p$.

\end{enumerate}
Thus, in this step, both $S^*$ (after swapping $p$ and $q$) and $\tilde S$ schedule the same packet $p$. Lemma~\ref{lemma:ops} is proved. $\blacksquare$
\end{proof}

Now, we present the algorithm {\tt ME}. {\tt ME} consists of maintaining packets in the queue (including selecting packets and updating their virtual deadlines) and delivering a packet at the end of each time step. For each new arrival $p$, its virtual deadline $t_p$ is initialized as its real deadline $d_p$. If there are more than one packet arriving, we consider them one by one. The deadline $d_p$ is specified by the adversary at its arrival. Then we calculate all existing packets in the queue and $p$ to find the optimal provisional schedule from time $t$. After we obtain the optimal provisional schedule, we update some packets' virtual deadlines, if necessary. Each packet updates its virtual deadline to the tentative time step specified in the optimal provisional schedule. At last, we send either the packet with the earliest virtual deadline (if it has a sufficient large value), or the maximum-value packet (otherwise). {\tt ME} is described in Algorithm~\ref{alg:me}.

\begin{algorithm}
\caption{{\tt ME}({\bf S}, \ t)}
\begin{algorithmic}[1]

\STATE For each new arrival $p$, set $t_p = d_p$.

\STATE Calculate the optimal provisional schedule $S^*$ by running ${\tt OPS}(Q^{\tt ME}_t \cup p, \ t)$.

\STATE Drop all packets not in $S^*$.

\STATE Update the virtual deadline $t_j$ of a packet $j \in S^*$ as $t + i$ ($\le d_j$), where $i$ is the index of the buffer slot that $j$ is residing in the optimal provisional schedule queue.

\COMMENT{Notice that $Q^{\tt ME}_t(i) = j$.}

\COMMENT{Let the packet with the earliest virtual deadline be $e$, let the maximum-value packet be $h$, with ties broken in favor of the earliest virtual deadline.}

\IF{$w_e \ge w_h / \alpha$}

\STATE Send $e$.

\ELSE

\STATE Send $h$.

\ENDIF

\COMMENT{Lines $5$ to $9$ are as in ${\tt EDF}_{\alpha}$~\cite{CJST07}.}

\end{algorithmic}

\label{alg:me}

\end{algorithm}

Directly from the algorithm {\tt ME}, for each $p \in Q^{\tt ME}_t$, we conclude the following properties of its virtual deadline $t_p$:
\begin{remark}
All packets $p \in Q^{\tt ME}_t$ have their $t_p$ sorted in strictly increasing order as $t, \ t + 1, \ \ldots, \ t + |Q^{\tt ME}_t| - 1$, where $|Q^{\tt ME}_t|$ is the number of packets in the queue. So, unless a new arrival $p$ comes with its virtual deadline $t_p = d_p > t + |Q^{\tt ME}_t| - 1$, accepting $p$ will lead to dropping exactly one packet in $Q^{\tt ME}_t$.
\label{remark:1}
\end{remark}
\begin{remark}
Every time when $t_p$ is updated (if any), $t_p$ is decreased strictly. For any packet $p \in Q^{\tt ME}_t$, $r_p \le t_p \le d_p$.
\label{remark:2}
\end{remark}
\begin{remark}
All packets in the buffer have distinct virtual deadlines, which may not be the same as their deadlines specified in the input sequence.
\label{remark:diff}
\end{remark}


\subsection{Analysis of {\tt ME}.}
\label{sec:analysisME}

\begin{theorem}
{\tt ME} is a deterministic $3$-competitive algorithm for scheduling packets with deadlines in the finite queue model, where the parameter $\alpha$ in the algorithm is set $2$.
\label{theorem:me}
\end{theorem}

Fix an input sequence of arriving packets. The actions of the algorithm can be regarded as a sequence of {\em packet arrival events} and {\em packet delivery events} $\tau := \tau_1 \tau_2 \ldots$. Then, in our algorithm {\tt ME} and its analysis, if not mentioning, we use the subscript $t$ to denote the event $\tau_t$, instead of the time step $t$. A single time step may involve more than one arrival events and only one delivery event.

In analyzing online algorithms, potential function approach and charging scheme are two commonly used methods~\cite{BY98}. The potential function method assigns some values as the potentials to the online algorithm and the adversary's configurations respectively, and then compares the change of the potentials in each time step to bound the competitive ratio. In our analysis, we use a modified potential function to prove Theorem~\ref{theorem:me}. We let {\tt ADV} denote the adversary of {\tt ME} and $\mathbb O$ denote the set of packets sent by {\tt ADV}, i.e., the packets in the optimal solution. Without loss of generality, we assume {\tt ADV} sends the earliest deadline packet in each time step. Let the packets sent by {\tt ADV} be $p_1, \ p_2, \ \ldots, \ p_i, \ \ldots$ in order. A packet $p_i \in \mathbb O$ is delivered in step $i$. If there is no packet to send in step $t$, $p_t$ is a {\em null} packet. Our analysis (especially for packet arrivals) depends on a critical observation on the adversary and a property of {\tt ME}:
\begin{remark}
Assume in steps $1, \ 2, \ \ldots, \ i, \ n$, {\tt ADV} sends packets $p_1, \ p_2, \ \ldots, \ p_i, \ \ldots, \ p_n$ in order. Clearly, $r_{p_i} \le i \le d_{p_i}$. Furthermore, in {\tt ADV}'s queue, we are free to modify the packets' deadlines $d_{p_i}$ to $t'_{p_i}$ as long as $r_{p_i} \le i \le t'_{p_i}$.
\end{remark}
\begin{remark}
For any packet $j$ in {\tt ME}'s queue, the minimum value of a packet $i$ with $t_i \le t_j$ does not decrease over time.
\label{remark:4}
\end{remark}

We use $\Phi^{\tt ME}_t$ (respectively, $\Phi^{\tt ADV}_t$) to denote the potential of the queue of {\tt ME} (respectively, {\tt ADV}) at time $t$. $\Phi^{\tt ME}_t$ (respectively, $\Phi^{\tt ADV}_t$) is the sum of the (mapped) $\mathbb O$-packets (respectively, $\mathbb O$-packets) in the queue. Define $S^{\tt A}_t$ as the packet sent by an algorithm {\tt A}. Our goal is to prove that at the end of each event, the following inequality
\begin{equation}
3 \cdot \sum_{j \in S^{\tt ME}_t} w_j + \Phi^{\tt ME}_t \ge \sum_{k \in S^{\tt ADV}_t} w_k + \Phi^{\tt ADV}_t,
\label{equ:main}
\end{equation}
holds. As a consequence, this yields Theorem~\ref{theorem:me}.

Let $\Delta^{\tt ME}_t$ (respectively, $\Delta^{\tt ADV}_t$) denote the difference of the left (respectively, right) side of Inequality~\ref{equ:main} from time $t - 1$ to time $t$, i.e.,
\begin{eqnarray}
\label{equ:mp}
\Delta^{\tt ME}_t & := & 3 \cdot \sum_{i \in (S^{\tt ME}_t \setminus S^{\tt ME}_{t - 1})} w_i + \Phi^{\tt ME}_t - \Phi^{\tt ME}_{t - 1}, \\
\label{equ:on}
\Delta^{\tt ADV}_t & := & \sum_{k \in (S^{\tt ADV}_t \setminus S^{\tt ADV}_{t - 1})} w_k + \Phi^{\tt ADV}_t - \Phi^{\tt ADV}_{t - 1}.
\end{eqnarray}

Obviously, Inequality~\ref{equ:main} holds before the first event since packets have not been sent so far. In order to prove Theorem 1, it is sufficient to prove that for each event, the following inequality holds since it leads to Inequality~\ref{equ:main}.

\begin{equation}
\Delta^{\tt ME}_t \ge \Delta^{\tt ADV}_t.
\label{equ:compare}
\end{equation}

In order to prove Inequality~\ref{equ:compare}, we present a set of invariants which hold at the end of each event.

\begin{itemize}

\item[$I_1.$]  $\Delta^{\tt ME}_t \ge \Delta^{\tt ADV}_t$.

\item[$I_2.$] {\tt ADV}'s queue contains only the set of packets it will send. For each packet $j \in (Q^{\tt ME}_t \cap Q^{\tt ADV}_t)$, {\tt ADV} has the virtual deadline $t_j$ as its real deadline.

For each packet $j \in Q^{\tt ME}_t$, $j$ maps to at most one packet $j' \in (Q^{\tt ADV}_t \setminus Q^{\tt ME}_t)$. For each packet $p' \in (Q^{\tt ADV}_t \setminus Q^{\tt ME}_t)$, $p'$ must be mapped uniquely by a packet $p \in Q^{\tt ME}_t$.

\item[$I_3$.] If $j \in Q^{\tt ME}_t$ maps to $j' \in (Q^{\tt ADV}_t \setminus Q^{\tt ME}_t)$, for any packet $i \in Q^{\tt ME}_t$ with $t_i \le t_j$, the following inequalities are true: $t_i \le d_{j'}$ and $w_i \ge w_{j'}$.

\end{itemize}

We prove that the set of invariants hold separately for both the events of packet arrivals and packet deliveries. Summing these inequalities over the arrivals and deliveries happened in one single time step yields the claim for a single time step; summing over all time steps proves Theorem~\ref{theorem:me}. To prove the existence of the above set of invariants, we apply case study in the following.

\begin{proof}
We show the set of invariants hold at the end of each time step. For each time step, we consider packet arrivals and packet delivery separately. In the following case analysis, we update the packets in {\tt ADV}'s queue as well as their mappings to the packets in {\tt ME}'s buffer.

If not mentioned otherwise, everything else remains unchanged at the end of this event. For ease of presentation, we assume buffer slots are indexed as $1, \ 2, \ \ldots, \ b$. We use $\mathbb O$ to denote the set of packets sent by the adversary. Let $e$ and $h$ denote the packets with the earliest virtual deadline and the packet with the maximum value, with ties broken in favor of the earliest virtual deadline one. Remember all packets in the queue have distinct virtual deadlines (see Algorithm~\ref{alg:ops} and Remark~\ref{remark:diff}).

We are going to show that in each time step, the ratio of {\tt ADV}'s gain over {\tt ME}'s gain is bounded by $1 + \alpha$, or $1 + 3 / \alpha$, or $2 + 2 / \alpha$. The competitive ratio $3$ is optimized at $\alpha = 2$ for
\begin{equation}
\min \max\{1 + \alpha, \ 1 + 3 / \alpha, \ 2 + 2 / \alpha\}.
\end{equation}


\subsubsection{Packet delivery}

In each time step, {\tt ME} either sends $e$ or $h$. If {\tt ME} sends $e$, $w_e \ge w_h / \alpha$; otherwise, {\tt ME} sends $h$. We assume {\tt ADV} sends $j$. At the end of this delivery event, $e$ is out of {\tt ME}'s queue because of its virtual deadline. We summarize all the possible consequences into the following $5$ cases, based on the packet {\tt ME} sends and the packet {\tt ADV} sends in each step.


\begin{enumerate}

\item Assume {\tt ME} and {\tt ADV} send the same packet $j$.

We charge {\tt ME} $w_j$. We charge {\tt ADV} $w_j$ initially. If $j \neq e$, i.e., $j = h$ and we know $w_h > \alpha \cdot w_e$. We charge {\tt ADV} more $w_e + w_{e'} + w_{j'}$, where $e'$ is the packet mapped by $e$ and $j'$ is the packet mapped by $j$, if any. From the invariant $I_3$, $\max\{w_{e'}, \ w_{j'}\} \le w_e$. Thus, the ratio of the modified gain for {\tt ADV} and {\tt ME} is bounded by $\max\{(w_e + w_{e'}) / w_e, \ (w_e + w_{e'} + w_{j} + w_{j'}) / w_j\} = \max\{2, \ 1 + 3 / \alpha\} = 1 + 3 / \alpha$, where $\alpha = 2$.


\item Assume {\tt ME} sends $e$ and {\tt ADV} sends $j \notin Q^{\tt ME}_t$.

From the invariant $I_2$, we assume $j = p'$, and $j$ is mapped by a packet $p \in Q^{\tt ME}_t$. Since {\tt ME} sends $e$, $w_e \ge w_h / \alpha$.

\begin{enumerate}

\item Assume $e = p$.

We charge {\tt ME} $w_e$ and charge {\tt ADV} $w_j \le w_p = w_e$. Then the ratio of the modified gains is $w_j / w_e \le 1$.

\item Assume $e \neq p$ and $e$ is not in any mapping.

We charge {\tt ME} $w_e$ and we {\tt ADV} $w_e + w_j$ (given $e$ possibly being an $\mathbb O$-packet). Then the ratio of the modified gains is $(w_e + w_j) / w_e \le (w_e + w_h) / w_e \le 1 + \alpha$.

\item Assume $e \neq p$ and $e$ maps $e' \in Q^{\tt ADV}_t$.

We have $d_{e'} > d_j$, otherwise, {\tt ADV} will select $e'$ to send (because we assume {\tt ADV} selects packets to send in the earliest deadline order). Thus, $d_{e'} \ge d_j = d_{p'} \ge t_p$ (the third inequality holds because of the invariant $I_2$). Also, $p \notin Q^{\tt ADV}$, otherwise, since $t_p = d_p \le d_j$, {\tt ADV} will send $p$ instead of $j$.

We charge {\tt ME} $w_e$ and we charge {\tt ADV} $w_{e'} + w_j$. Then the ratio of the modified gains is $(w_{e'} + w_j) / w_e \le (w_e + w_h) / w_e = 1 + \alpha$.

\end{enumerate}


\item Assume {\tt ME} sends $e$ and {\tt ADV} sends $j \in Q^{\tt ME}_t$, $e \neq j$.

$w_e \ge w_h / \alpha \ge w_j / \alpha$. $e \notin Q^{\tt ADV}_t$, otherwise, {\tt ADV} will send $e$ instead of $j$ in this step.

We charge {\tt ME} $w_e$. We charge {\tt ADV} $w_{e'} + w_j$, assuming $e$ maps $e'$. Then the ratio of the modified gains is $(w_{e'} + w_j) / w_e \le 1 + \alpha$. If $j$ maps a packet in {\tt ADV}'s queue only, this mapping still holds at the end of this delivery event.


\item Assume {\tt ME} sends $h \neq e$ and {\tt ADV} sends $j \notin Q^{\tt ME}_t$.

Note $w_e < w_h / \alpha$. From the invariant $I_2$, we can assume $p \in Q^{\tt ME}_t$ maps $p' = j \in (Q^{\tt ADV}_t \setminus Q^{\tt ME}_t)$. Since {\tt ADV} sends $p'$ and $d_{p'} \ge t_p$ (from the invariant $I_2$), we know $p \notin Q^{\tt ADV}_t$. Thus, $w_j = w_{p'} \le w_e$. $h$ should be in $Q^{\tt ADV}_t$, otherwise, {\tt ADV} can send $h$ instead of $j$ to gain more value in this time step. Also, $w_{h'} \le w_e$.

\begin{enumerate}

\item Assume $e = p$.

$e$ is out of {\tt ME}'s queue at the end of this event because of its virtual deadline expires. Then we charge {\tt ME} $w_h$. We charge {\tt ADV} $w_e + w_{e'} + w_h + w_{h'}$, assuming $e$ maps $e' = j$ and $h$ maps $h'$. Note $\max\{w_{e'}, \ w_{h'}\} \le w_e < w_h / \alpha$. Then the ratio of the modified gains is $(w_e + w_{e'} + w_h + w_{h'}) / w_h \le (3 + \alpha) / \alpha = 1 + 3 / \alpha$.

\item Assume $e \neq p$.

Assume $e \neq p$ and $e$ maps $e' \in Q^{\tt ADV}_t$. We have $d_{e'} > d_j$, otherwise, {\tt ADV} will select $e'$ to send (because we assume {\tt ADV} selects packets to send in earliest deadline order). Thus, $d_{e'} \ge d_j = d_{p'} \ge t_p$.

We map $p$ to $e'$. We charge {\tt ME} $w_h$ and we charge {\tt ADV} $w_h + w_{h'} + w_e + w_{e'}$. Then the ratio of the modified gains is $(w_e + w_{e'} + w_h + w_{h'}) / w_h \le 1 + 3 / \alpha$.

\end{enumerate}


\item Assume {\tt ME} sends $h \neq e$ and {\tt ADV} sends $j \in Q^{\tt ME}_t$, $j \neq h$.

Clearly, $t_j = d_j < t_h = d_h$, otherwise, {\tt ADV} can always swap the sending sequences of $j$ and $h$. $h$ should be in $Q^{\tt ADV}_t$, otherwise, {\tt ADV} can send $h$ instead of $j$ to gain more value in this time step. Also, $w_{h'} \le w_e$.

\begin{enumerate}

\item Assume $j = e$.

We charge {\tt ME} $w_h$ and charge {\tt ADV} $w_h + w_{h'} + w_e + w_{e'}$, assuming $e$ maps $e'$ and $h$ maps $h'$. $\max\{w_{e'}, \ w_{h'}\} \le w_e \le w_h / \alpha$. Then the ratio of the modified gains is $(w_h + w_{h'} + w_e + w_{e'}) / w_h \le 1 + 3 / \alpha$.

\item Assume $j \neq e$.

$e$ is not an $\mathbb O$-packet. Assume $e$ maps $e'$ in {\tt ADV}'s queue; if $e'$ does not exist, we let $e'$ be a null packet. $w_{e'} \le w_e$. Note $w_{j'} \le w_e$. At the end of this delivery, $j$ is still in {\tt ME}'s queue but $j$ is not in {\tt ADV}'s queue.

We charge {\tt ME} $w_h$. Then, we charge {\tt ADV} $w_h + w_{h'} + w_j + w_{e'}$. Then the ratio of the modified gains is $(w_h + w_{h'} + w_j + w_{e'}) / w_h \le (w_h + w_e + w_h + w_e) / w_h = 2 + 2 / \alpha$.

\end{enumerate}

\end{enumerate}


\subsubsection{Packet arrivals}

Remember that from the properties of algorithm {\tt ME} (see Remark~\ref{remark:1} and Remark~\ref{remark:2}), for each new arrival $p$, if admitting $p$ results in a packet $i$ leaving $Q^{\tt ME}_t$, the total value of the queue is not decreasing. $p$ can always be a candidate packet to map the packet which was mapped by the packet evicted due to accepting $p$ (since $w_p \ge w_i$). Also, the slack time of $p$, defined as $d_p - t$, is no larger than the total number of packets in the queue $|Q^{\tt ME}_t|$, otherwise, $p$ will be accepted by {\tt ME} without evicting a packet (see Algorithm~\ref{alg:ops}).

For each arriving event at time $t$, $S^{\tt ME}_{t + 1} \setminus S^{\tt ME}_t = Q^{\tt ADV}_{t + 1} \setminus Q^{\tt ADV}_t = \emptyset$. Thus, for each new arrival event, we only need to consider the change of mappings, if any.

From the property of the adversary, we know that all packets in {\tt ADV}'s queue (supposed to be sent by {\tt ADV}) will not be evicted when we put new $\mathbb O$-packets in {\tt ADV}'s queue. Let us consider the case when introducing an $\mathbb O$-packet $p$ results in a packet $i$ leaving {\tt ME}'s queue. If $i$ is not in {\tt ADV}'s queue, we are fine with all mappings and potentials because there is no loss to $\Phi^{\tt ME}_t$. We only consider the case when $i$ is in {\tt ADV}'s queue.

Assuming $i$ is an $\mathbb O$-packet in {\tt ADV}'s queue, we first claim that we can always find a packet $q$, which is not in {\tt ADV}'s queue with $w_q \ge w_i$. Otherwise, {\tt ADV} does not accept $p$ as well. Then we collect all $\mathbb O$-packets in {\tt ME}'s queue but with deadlines $\le |Q^{\tt ME}_t|$, as they are need to be delivered by {\tt ADV} by time $t + |Q^{\tt ME}_t|$, it does not hurt to assign them deadlines in strictly decreasing order from $t + |Q^{\tt ME}_t|$.  Therefore, the evicted $\mathbb O$-packet $i$ can be assigned a deadline as the virtual deadline of the latest non-$\mathbb O$-packet in a buffer slot no later than $d_i - t$ in {\tt ME}'s queue. Let this packet be $j$. We can remove $i$ from {\tt ADV}'s queue and put $j$ with $t_j$ as its deadline and $w_j$ as its value in {\tt ADV}'s queue. These operations do not hurt {\tt ADV} because of Remark~\ref{remark:4}. Above reasoning can also be applied to the case when a new $\mathbb O$-packet $p$ is rejected by {\tt ME}.

Based on our case study at packet arrival events and packet delivery events discussed above, Theorem~\ref{theorem:me} is proved. $\blacksquare$
\end{proof}

Some side research results on the finite queue model are shown as follows; they give the upper and lower bounds that (some) online algorithms can achieve.

We use ${\bf S}_t$ to denote both the provisional schedule for time steps $[t, \ +\infty)$ and the set of packets specified by the schedule. All known online algorithms for the bounded-delay model~\cite{KLMPSS04}, \cite{H01}, \cite{CF03}, \cite{LSS05}, \cite{LSS07}, \cite{EW07} calculate their optimal provisional schedules at the beginning of each time step. These algorithms only differ by the packets they select to send. The deterministic online algorithms in such a broad family are defined as the {\em best-effort admission algorithms}.

\begin{definition}
{\bf Best-effort admission algorithm}. Consider an online algorithm {\tt ON} and a set of pending packets ${\bf P}_t$ at time $t$. If {\tt ON} calculates the optimal provisional schedule ${\bf S}_t$ on ${\bf P}_t$ and selects one packet from ${\bf S}_t$ to send in the step $t$, we call {\tt ON} a {\em best-effort admission algorithm}.
\end{definition}

\begin{theorem}
The lower bound of competitive ratio for the best-effort admission algorithms is $2$.
\label{theorem:lowerbound}
\end{theorem}

\begin{proof}
In the following instance, we will show: {\em If the buffer size is bounded, the packets that the optimal offline algorithm chooses to send may not be from the optimal provisional schedule calculated by the online algorithm, even if both algorithms have the same set of pending packets}. This property does not hold in the bounded-delay model; and it leads that any deterministic best-effort admission algorithm cannot achieve a competitive ratio better than $2$.

Assume the buffer size is $b$. Let a best-effort admission online algorithm be {\tt ON}. We use $(w_p, \ d_p)$ to represent a packet $p$ with a value $w_p$ and a deadline $d_p$. Initially, the buffer is empty.  A set of packets, from which the optimal offline algorithm will accept $b - 1$ packets from them and eventually send, are released: $(1, \ b + 1), \ (1, \ b + 2), \ \ldots, \ (1, \ b + b)$. Based on its definition, upon these packets' arrival, {\tt ON} will select all of them to put into its buffer. Notice that all packets released have deadlines larger than the buffer size $b$. The optimal offline algorithm drops $(1, \ b + 1)$, and keeps $(1, \ b + 2), \ \ldots, \ (1, \ b + b)$ in its buffer.

In the same time step, $b$ packets $(1 + \epsilon, \ 1), \ (1 + \epsilon, \ 2), \ \ldots, \ (1 + \epsilon, \ b)$ are released afterwards. There are no more new packets arriving in this step. The optimal offline algorithm only accepts $(1 + \epsilon, \ 1)$. Thus, after processing arrivals in step $1$, the optimal offline algorithm send the packet $(1 + \epsilon, \ 1)$. Instead, {\tt ON} calculates the optimal provisional schedule in step $1$ which includes all these newly arriving packets with value $1 + \epsilon$. All such packets will be accepted by {\tt ON}, but the packets $(1, \ b + i)$, $\forall i = 1, \ 2, \ \ldots, \ b$, will be dropped due to the buffer size constraint. {\tt ON} sends a packet with value $1 + \epsilon$ in the first step.

At the beginning of each step $i = 2, \ 3, \ \ldots, \ b$, only one packet $(1 + \epsilon, \ i)$ is released. At the end of step $b$, no new packets will be released in the future. Since the time after the first step, all packets available to {\tt ON} have their deadlines $\le b$. Thus, {\tt ON} cannot schedule sending packets with a total value $\ge (1 + \epsilon) \cdot (b - 1)$ in the time steps $2, \ 3, \ \ldots, \ b$.

Of course, since there is one empty buffer slot at the beginning of each time step $i = 2, \ 3, \ \ldots, \ b$, the optimal offline algorithm can accept and send all newly released packets $(1 + \epsilon, \ i)$ in steps $i = 2, \ 3, \ \ldots, \ b$. At the end of step $b$, the packets $(1, \ b + 2), \ (1, \ b + 3), \ \ldots, \ (1, \ b + b)$ are still remained in the optimal offline algorithm's buffer (they are not in {\tt ON}'s buffer though). Since there is no future arrivals, these $b - 1$ packets will be transmitted eventually by the optimal algorithm in the following $b - 1$ steps. The total value of {\tt ON} achieves is $(1 + \epsilon) \cdot b$ while the optimal offline algorithm gets a total value $(1 + \epsilon) \cdot b + 1 \cdot (b - 1)$. The competitive ratio for this instance is
\begin{displaymath}
c = \frac{(1 + \epsilon) \cdot b + 1 \cdot (b - 1)}{(1 + \epsilon) \cdot b} = 2 - \frac{1 + b \cdot \epsilon}{b + b \cdot \epsilon} \ge 2 - \frac{2}{b}, \ \ \ \mbox{if } \epsilon \cdot b = 1 \mbox{ and } b \ge 2.
\end{displaymath}

If $b$ is large, {\tt ON} cannot perform asymptotically better than $2$-competitive. This lose is due to {\tt ON} calculating optimal provisional schedule to find out the packet to send in each time step. Theorem~\ref{theorem:lowerbound} is proved. $\blacksquare$
\end{proof}

\begin{lemma}
The simple greedy algorithm, which selects packets in the optimal provisional schedule and sends the maximum-value packet in each time step, is no better than $4$-competitive.
\label{lemma:greedy}
\end{lemma}

\begin{proof}
In the following instance, we will show that the greedy algorithm {\tt Greedy}, which calculates the optimal provisional schedule and schedules the maximum-value packet, cannot be better than $4$-competitive. Let the buffer size be $b$. Without loss of generality, we assume $b$ is even. We use $(w_p, \ d_p)$ to represent a packet $p$ with a value $w_p$ and a deadline $d_p$.

Suppose at the end of step $0$, the buffer is empty.  A set of packets, which the optimal offline algorithm will eventually send, are released: $(1, \ b + 1), \ (1, \ b + 2), \ \ldots, \ (1, \ b + b)$. Notice that all packets in the buffer have deadlines larger than the buffer size $b$.

At the beginning of step $1$, $b$ packets $(1 + 1 \cdot \epsilon, \ 1), \ (1 + 2 \cdot \epsilon, \ 2), \ \ldots, \ (1 + b \cdot \epsilon, \ b)$ are released. {\tt Greedy} accepts all these newly arriving packets. The optimal offline algorithm only accepts $(1 + \epsilon, \ 1)$, drops $(1, \ b + 1)$, and keeps $(1, \ b + 2), \ \ldots, \ (1, \ b + b)$ in its buffer. In step $1$, the optimal offline algorithm send $(1 + \epsilon, \ 1)$. Instead, {\tt Greedy} will accept all newly released packets in step $1$, thus, all packets $(1, \ b + i)$ for any $i \ = \ 1, \ 2, \ \ldots, \ b$ are dropped. {\tt Greedy} sends the packet $(1 + b \cdot \epsilon, \ b)$. At the end of this step, the packet $(1 + 1 \cdot \epsilon, \ 1)$ in {\tt Greedy}'s buffer expires.

At the beginning of each step $i \ = \ 2, \ 3, \ \ldots, \ b$, only one packet $(1 + \epsilon, \ i)$ is released. At the end of step $b$, no future packets will be released. {\tt Greedy} rejects all these newly released packets. {\tt Greedy} will send the packets $(1 + (b - 1) \cdot \epsilon, \ b - 1), (1 + (b - 2) \cdot \epsilon, \ b - 2), \ \ldots, \ (1 + (b / 2 + 1) \cdot \epsilon, \ b / 2 + 1)$ in the following $b / 2 - 1$ time steps. All the packets $(1 + 2 \cdot \epsilon, \ 2), \ (1 + 3 \cdot \epsilon, \ 3), \ \ldots, \ (1 + (b / 2) \cdot \epsilon, \ b / 2)$ will be dropped due to their deadlines.

Of course, the optimal offline algorithm can send all newly released packets in steps $2, \ 3, \ \ldots, \ b$. At the end of step $b$, the packets $(1, \ b + 2), \ (1, \ b + 3), \ \ldots, \ (1, \ b + b)$ are still remained in the optimal offline algorithm's buffer, but not in {\tt Greedy}'s buffer. Since there is no future arrivals, these $b - 1$ packets will be transmitted eventually by the optimal algorithm in the following $b - 1$ steps. If $b$ is large, {\tt Greedy} cannot perform better than $4$-competitive. The lose is due to {\tt Greedy}'s first step in which optimal provisional schedule is used in selecting packets in the buffer and its greedy manner in sending packets in the first $b / 2$ time steps. Lemma~\ref{lemma:greedy} is proved. $\blacksquare$
\end{proof}


\section{Algorithm {\tt RME} and Its Analysis}
\label{sec:rme}

In this section, we present a randomized algorithm for the finite capacity queue model. Our algorithm is named {\tt RME}, which stands for ``Randomized {\tt ME}''. Similar to {\tt ME}, {\tt RME} consists of two parts in handling packet arrivals and packet deliveries respectively in each time step. The difference is that {\tt RME} employs a random variable to decide whether to send the earliest packet or the most valuable packet.

We would like to point out that even {\tt RME} makes a random choice during its execution, this randomization is executed by the algorithm internally and has nothing to do with the characteristics of the input sequence. The adversary is allowed to generate the input sequence to maximize the competitive ratio. No stochastic assumption is made on the input sequence.


\subsection{Algorithm {\tt RME}.}

For each packet arrival event, {\tt RME} works the same as what {\tt ME} does. That is, {\tt RME} calls ${\tt OPS({\bf S}, t)}$ to identify the packets in its buffer deterministically, where ${\bf S}$ is the set of pending packets and $t$ is the current time. In packet delivery, a random variable $\beta$ is used to facilitate scheduling.

We use $e$ to denote the packet with the earliest-virtual deadline packet and $h$ to denote the earliest maximum value packet in the buffer. The algorithm works as follows. If $e$ has a sufficiently large value with $w_e \ge w_h / \alpha$, we send $e$ deterministically. Otherwise (i.e., $w_e < w_h / \alpha$), we choose $\beta$ uniformly on $[0, \ 1]$. If $\beta \in [0, \ \gamma]$ (we will decide $\gamma$ later. The parameter $\gamma$ influences the competitive ratio of the algorithm), we deliver $e$, otherwise (i.e., if $\beta \in (\gamma, \ 1]$ and $w_e < w_h / \alpha$), we deliver $h$. The pseudo code of {\tt RME} is described in Algorithm~\ref{alg:rme}, where the set of pending packets at time $t$ is ${\bf S}$.

\begin{algorithm}
\caption{{\tt RME}({\bf S}, \ t)}
\begin{algorithmic}[1]

\STATE For each new arrival $p$, set $t_p = d_p$.

\STATE Calculate the optimal provisional schedule $S^*$ by running ${\tt OPS}(Q^{\tt ME}_t \cup p, \ t)$.

\STATE Drop all packets not in $S^*$.

\STATE Update the virtual deadline $t_j$ of a packet $j \in S^*$ as $t + i$, where $i$ is the index of the buffer slot that $j$ is in.

\IF{$w_e \ge w_h / \alpha$}

\STATE Send $e$.

\ELSE

\STATE Choose $\beta$ uniformly on $[0, \ 1]$.

\IF{$\beta \in [0, \ \gamma]$}

\STATE Send $e$.

\ELSE

\STATE Send $h$.

\ENDIF

\ENDIF

\end{algorithmic}
\label{alg:rme}
\end{algorithm}


\subsection{Analysis of {\tt RME}.}

\begin{theorem}
{\tt RME} is a randomized ($\phi^2 \approx 2.618$)-competitive algorithm for scheduling packets with deadlines in the finite queue model, where $\alpha = \phi \approx 1.618$ and $\gamma = 1 / \phi^2 \approx 0.382$.
\label{theorem:rme}
\end{theorem}

Some formulas are used in our analysis:
\begin{displaymath}
1 / \phi + 1 = \phi, \ \ \phi + 1 = \phi^2, \ \ 2 + 1 / \phi = \phi^2, \ \ \phi + 1 / \phi^2 = 2.
\end{displaymath}

\begin{proof}
At first, we examine the expected gain that the algorithm {\tt RME} can gain in a time step. From the algorithm itself, we know that {\tt RME} gains,
\begin{eqnarray*}
w_e, & & \mbox{ if } w_e \ge w_h / \alpha, \\
w_e, & & \mbox{ if } w_e < w_h / \alpha \mbox{ and } \beta \in [0, \ 1 / \phi^2], \\
w_h, & & \mbox{ if } w_e < w_h / \alpha \mbox{ and } \beta \in (1 / \phi^2, \ 1].
\end{eqnarray*}

In either way ($w_e \ge w_h / \phi$ or $w_e < w_h / \phi$), the algorithm gains an expected value of
\begin{displaymath}
{\bf E}(W_t) \ \ge \ \min\{w_h / \phi, \ w_e \cdot (1 / \phi^2) + w_h \cdot (1 - 1 / \phi^2)\} \ = \ w_h / \phi.
\end{displaymath}

Remember $S^{\tt A}_t$ is the packet sent by an algorithm {\tt A}. We observe the following set of invariants about {\tt ADV} and {\tt RME}'s buffers:
\begin{itemize}

\item[$V_1$.]
\begin{displaymath}
\phi^2 \cdot {\bf E}(\sum_{j \in S^{\tt RME}_t} w_j) + \Phi^{\tt RME}_t = \phi^2 \cdot \sum_{j \in S^{\tt RME}_t} {\bf E}(w_j) \ge \sum_{k \in S^{\tt ADV}_t} w_k + \Phi^{\tt ADV}_t,
\end{displaymath}
where linearity of expectations is used in the first equality.

\item[$V_2$.] {\tt ADV}'s queue contains only the set of packets it sends. For each packet $j \in (Q^{\tt RME}_t \cap Q^{\tt ADV}_t)$, {\tt ADV} has the virtual deadline $t_j$ as this packet's modified real deadline $d_j$.

For each packet $j \in Q^{\tt RME}_t$, $j$ maps to at most one packet $j' \in (Q^{\tt ADV}_t \setminus Q^{\tt RME}_t)$. For each packet $p' \in (Q^{\tt ADV}_t \setminus Q^{\tt RME}_t)$, $p'$ must be mapped uniquely by a packet $p \in Q^{\tt RME}_t$.

\item[$V_3$.] If $j \in Q^{\tt RME}_t$ maps $j' \in (Q^{\tt ADV}_t \setminus Q^{\tt RME}_t)$, for any packet $i \in Q^{\tt RME}_t$ with $t_i \le t_j$, $t_i \le d_{j'}$ and $w_i \ge w_{j'}$.

\end{itemize}

Similar to the analysis of {\tt ME}, packet arrival and delivery events are analyzed separated from case studies. We omit the analysis on packet arrival, which has been presented in the proof of Theorem~\ref{theorem:me}. In the following, we discuss the invariants in the randomized packet delivery events.

In each time step, {\tt RME} either sends $e$ or $h$. If {\tt RME} sends $h$, we must have $w_h \ge \alpha \cdot w_e = \phi \cdot w_e$ (from the algorithm). We assume {\tt ADV} sends $j$. At the end of this delivery event, $e$ is out of {\tt RME}'s queue because of its virtual deadline. We study the following cases which are categorized based on the packet {\tt RME} sends and the packet {\tt ADV} sends in each time step.


\begin{enumerate}

\item Assume {\tt RME} sends $e$ and $w_e \ge w_h / \phi$.

As what we have seen in the proof of Theorem~\ref{theorem:me}, the ratio of the modified gain is bounded by $1 + \alpha = 1 + \phi = \phi^2$.

\item Assume {\tt RME} sends $e$ with $w_e < w_h / \phi$.

This case happens with a probability of $1 / \phi^2$ when $w_e < w_h / \phi$. We combine this case with the next case to get the expected competitive ratio.

\item Assume {\tt RME} sends $h \neq e$ and {\tt ADV} sends $j \neq h$.

This case happens with a probability of $1 / \phi$ when $w_e < w_h / \phi$. $e$ leaves the buffer at the end of this delivery due to its virtual deadline. We assume $j \neq e$.

(If $j = e$, we just ignore all $w_j$ and $w_{j'}$ in our following calculation and all inequalities still hold with simpler representations. If $j = h$, we simply use $h$ to replace $j$ in the following calculation and all inequalities hold also.)

We categorize the cases with $e$ is in a mapping and $e$ is not in a mapping.

If $e$ is in a mapping, $e$ is sent with a probability of $1 / \phi^2$ and $h$ is sent with a probability of $1 / \phi$. In the first case, we charge {\tt ADV} $w_e + w_{e'}$ where $e$ maps $e'$. In the second case, we charge {\tt ADV} $w_e + w_{e'} + w_j + w_{h'}$. Thus, the expected competitive ratio is
\begin{eqnarray*}
c & = & \frac{{\tt ADV}_t}{{\bf E}({\tt RME}_t)} \\ & = & \frac{(1 / \phi^2) \cdot (2 \cdot w_e) + (1 / \phi) \cdot (w_e + w_{e'} + w_j + w_{h'})}{(1 / \phi^2) \cdot w_e + (1 / \phi) \cdot w_h} \le \frac{2 w_e + \phi \cdot 2 w_e + \phi \cdot 2 w_h}{w_e + \phi \cdot w_h} \\ & \le & \frac{(2 \phi + 2) (w_h / \phi) + 2 \phi \cdot h}{w_h / \phi + \phi \cdot w_h} = \frac{2 + 2 / \phi + 2 \phi}{1 / \phi + \phi} \le \phi^2.
\end{eqnarray*}

If $e$ is not in a mapping, $e$ is sent with a probability of $1 / \phi^2$ and $h$ is sent with a probability of $1 / \phi$. In the first case, we charge {\tt ADV} $w_j \le w_h$. In the second case, we charge {\tt ADV} $w_j + w_h + w_{h'}$. Remember $w_e \ge w_{h'}$ (see the invariant $V_3$). Thus, the expected competitive ratio is
\begin{eqnarray*}
c & = & \frac{{\tt ADV}_t}{{\bf E}({\tt RME}_t)} \\ & = & \frac{(1 / \phi^2) \cdot w_j + (1 / \phi) \cdot (w_j + w_h + w_{h'})}{(1 / \phi^2) \cdot w_e + (1 / \phi) \cdot w_h} \le \frac{w_h / \phi^2 + (w_h + w_h + w_e) / \phi}{w_e / \phi^2 + w_h / \phi} \\ & \le & \frac{w_h + 2 \cdot \phi \cdot w_h + \phi \cdot w_e}{w_e + \phi \cdot w_h} \le \frac{2 \cdot \phi + 1}{\phi} = \phi^2.
\end{eqnarray*}
\end{enumerate}


Based on the above analysis, Theorem~\ref{theorem:rme} is proved. $\blacksquare$
\end{proof}


\section{The Optimal Offline Algorithm and Its Analysis}
\label{sec:offline}

Let $\mathbb O$ denote the set of packets sent by an optimal offline algorithm. Our algorithm is simple. Fix an input sequence $\cal I$. We start from a set of packets ${\bf S}_0 \subseteq {\cal I}$ such that all packets in ${\bf S}_0 = {\bf S}$ can be delivered successfully if we send them in increasing order of deadlines. Given a set of packets ${\bf S}$, we get the total gain of $W({\bf S})$. If ${\bf S}_0 = {\cal I}$, the algorithm is optimal. If ${\bf S}_0 \neq {\cal I}$, then we study the set of packets ${\cal I} \setminus {\bf S}_0$. We sort all packets in ${\cal I} \setminus {\bf S}_0$ in increasing order of deadlines. For each packet $j$, adding $j$ into ${\bf S}_0$ generates a new set ${\bf S}$; this results at most one packet $i$ not being sent successfully (still under the earliest deadline policy). Then we run into a loop to pick up a packet $i \in {\bf S}$ with $w_i < w_j$ and see whether $W({\bf S} \cup \{j\} \setminus \{i\}) > W({\bf S})$. ($i$ can be a {\em null packet} such that ${\bf S} \cup \{j\} \setminus \{i\} = {\bf S} \cup \{j\}$. If $W({\bf S} \cup \{j\}) > W({\bf S} \cup \{j\} \setminus \{i\}) > W({\bf S})$, we add $j$ into ${\bf S}$. If $W({\bf S} \cup \{j\}) < W({\bf S} \cup \{j\} \setminus \{i\})$, we drop $i$. We iteratively move $j$ out of ${\cal I} \setminus {\bf S}_0$ into ${\bf S}$ until ${\cal I} \setminus {\bf S}_0$ is empty. We claim that the schedule we finally have is optimal. It depends on the following two theorems.

\begin{theorem}
Given a set of packets ${\bf S}$, if all packets can be sent by their deadlines with the buffer size constraint, we can always schedule them in increasing order of deadlines among all pending packets in the buffer.
\end{theorem}

\begin{proof}
This is a standard result. $\blacksquare$
\end{proof}

\begin{lemma}
${\bf S}_0$ is easy to be construct. We simply pick up the earliest deadline packet to send and in each time step, greedily accept packets.
\end{lemma}

\begin{theorem}
If a set of packets ${\bf S}$ can be delivered successfully, ${\bf S} \cup \{j\}$ results at most one packet unsuccessfully sent. We can pick up any packet $i$ with $w_i < w_j$ as the candidate and schedule ${\bf S} \cup \{j\} \setminus \{i\}$. If $W({\bf S} \cup \{j\} \setminus \{i\}) > W({\bf S})$ but $W({\bf S} \cup \{j\}) \le W({\bf S})$, $i \notin {\mathbb O}$.
\end{theorem}

\begin{proof}
This is due to the property of matriod. $\blacksquare$
\end{proof}

\begin{theorem}
The optimal offline algorithm runs in polynomial-time $O(n^3)$, where $n$ is the size of the input sequence.
\end{theorem}

\begin{proof}
Given a set of $m$ packets ${\bf S}$, sorting all packets in ${\bf S}$ (in order of increasing deadlines) takes time $O(m \cdot \log m)$. For each packet not in ${\bf S}$ but not discarded yet, we take time $O(m)$ to locate a packet $i$ with value $w_i < w_j$ and then we schedule ${\bf S} \cup \{j\} \setminus \{i\}$ (note that $i$ can be a {\em null packets} such that ${\bf S} \cup \{j\} \setminus \{i\} = {\bf S} \cup \{j\}$). It takes time $O(m)$ to get the total gain of a schedule. Thus, identifying whether to accept $j$ or to discard $j$ takes time $O(m^2)$. Let the set of packets in the input sequence be $n$ with $n > m$. The total running time of our algorithm is bounded by $O(n \cdot n^2) = O(n^3)$. $\blacksquare$
\end{proof}


\section{Conclusions and Open Problems}
\label{sec:conclusion}

In this paper, we present two online algorithms for scheduling weighted packets with hard deadlines in a finite capacity queue and we provide their theoretical competitive analysis. The model we study generalizes the extensively studied bounded-delay model for QoS buffer management. Our model has significant importance in real system design since it takes into account of the realistic bound on the size of the router queues. The deterministic memoryless algorithm we present is $3$-competitive and the randomized memoryless algorithm we present is ($\phi^2 \approx 2.618$)-competitive. Both algorithms provide the worst-case guarantees to robustly optimize our objective (maximizing the weighted throughput) without applying any stochastic assumptions over the packet traffic. We propose a novel analysis approach by updating packets' parameters in an online manner. Instead of using the real deadlines, we introduce virtual deadlines, which are updated over time to help us make the best decision on when to send the packets. The virtual deadlines are strictly decreased over time and they guarantee the hard deadlines are always satisfied. This idea can be applied in many other online and real-time problems.

Closing or shrinking the gap of $[1.618, \ 2.618]$ between the lower bound and upper bound of competitive ratios for this model is an open problem. For a broad family of online algorithms, including all previously known research for the bounded-delay model, the lower bound is $2$. Our algorithmic framework can be applied to the multi-buffer model~\cite{AL06}. Getting an algorithm better than $9.82$-competitive for the the multi-buffer model is still an interesting open problem.


\bibliographystyle{plain}
\bibliography{buffer}


\end{document}